\documentclass{vgtc}                          % final (conference style)
\ifpdf%                                % if we use pdflatex
  \pdfoutput=1\relax                   % create PDFs from pdfLaTeX
  \usepackage{graphicx}                % allow us to embed graphics files
  \DeclareGraphicsExtensions{.pdf,.png,.jpg,.jpeg} % for pdflatex we expect .pdf, .png, or .jpg files
\else%                                 % else we use pure latex
  \usepackage{graphicx}                % allow us to embed graphics files
  \DeclareGraphicsExtensions{.eps}     % for pure latex we expect eps files
\fi%

\graphicspath{{figures/}{pictures/}{images/}{./}} % where to search for the images

\usepackage{microtype}                 % use micro-typography (slightly more compact, better to read)
\PassOptionsToPackage{warn}{textcomp}  % to address font issues with \textrightarrow
\usepackage{textcomp}                  % use better special symbols
\usepackage{mathptmx}                  % use matching math font
\usepackage{times}                     % we use Times as the main font
\usepackage{cite}                      % needed to automatically sort the references
\usepackage{tabu}                      % only used for the table example
\usepackage{booktabs}                  % only used for the table example
 \usepackage{multirow}
 \usepackage[table,xcdraw]{xcolor}
\usepackage{mathtools}
\providecommand{\etal}[0]{\emph{~et~al.}}

\providecommand{\vid}[0]{360\degree~video }
\providecommand{\vids}[0]{360\degree~videos }

\usepackage{gensymb}
\usepackage[acronym,shortcuts]{glossaries}

\newacronym{fov}{FOV}{field of view}
\newacronym{sota}{SotA}{state-of-the-art}
\newacronym{hmd}{HMD}{head-mounted display}
\newacronym{vr}{VR}{virtual reality}
\newacronym{mpeg}{MPEG}{Moving Picture Experts Group}
\newacronym{omaf}{OMAF}{omnidirectional media format}
\newacronym{mcts}{MCTS}{motion constrained tile sets}
\newacronym{obmc}{OBMC}{overlapped-block motion compensation}
\newacronym{dct}{DCT}{discrete cosine transform}
\newacronym{cdf}{CDF}{Cohen–Daubechies–Feauveau}
\newacronym{lgt}{LGT}{LeGall-Tabatabai}
\newacronym{fwt}{FWT}{fast wavelet transform}
\newacronym{ifwt}{iFWT}{inverse fast wavelet transform}
\newacronym{crf}{CRF}{constant rate factor}
\newacronym{fps}{FPS}{frames per second}
\newacronym{cgi}{CGI}{computer-generated imagery}
\newacronym{dof}{DOF}{degrees of freedom}
\usepackage{multirow} %% For Errors table
\usepackage{microtype}
\usepackage
[disable]
{todonotes}

\usepackage{pgfplots}
\onlineid{1590}

\vgtccategory{Technical papers}%{Research}

\vgtcinsertpkg

\title{Wavelet-Based Fast Decoding of 360\degree~Videos}

\author{Colin Groth, Sascha Fricke, Susana Castillo, and Marcus Magnor} %,~\IEEEmembership{Senior,~IEEE}}
\authorfooter{
\item
 All authors are with the Institute for Computer Graphics at TU Braunschweig, Germany. \\E-mail: \{groth,~fricke,~castillo,~magnor\}@cg.cs.tu-bs.de.
}

\teaser{
  \centering
  \includegraphics[width=\textwidth]{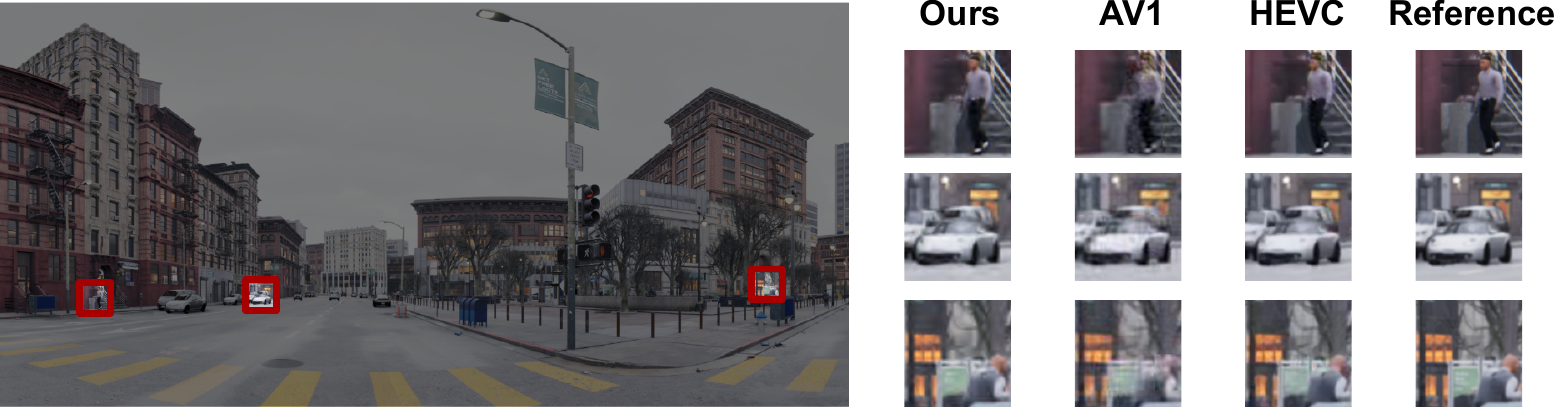}
  \caption{In this paper, we propose a wavelet-based video codec that is able to load and decode the data of \vids viewport-dependently by exploiting the properties of the wavelet transform. While the quality is on par with \acl{sota} video codecs, we can achieve significantly faster playback times. 
  Here, a visual comparison of our codec is given with HEVC, AV1 and the uncompressed reference. The samples are out of the computer-generated \textit{City} video (left), all with 8k full-frame resolution. %\cg{to compare the samples DOWNLOAD the pdf first. The latex previewer seems to apply a blur on the images too ..}
  }
  \label{fig:teaser}
}

\abstract{
In this paper, we propose a wavelet-based video codec specifically designed for VR displays that enables real-time playback of high-resolution 360\degree~videos. Our codec exploits the fact that only a fraction of the full 360\degree~video frame is visible on the display at any time.
To load and decode the video viewport-dependently in real time, we make use of the wavelet transform for intra- as well as inter-frame coding. Thereby, the relevant content is directly streamed from the drive, without the need to hold the entire frames in memory.
With an average of $193$ frames per second at $8192 \times 8192$ -pixel full-frame resolution, the conducted evaluation demonstrates that our codec's decoding performance is up to $272\%$ higher than that of the state-of-the-art video codecs H.265 and AV1 for typical VR displays. 
By means of a perceptual study, we further illustrate the necessity of high frame rates for a better VR experience.
Finally, we demonstrate how our wavelet-based codec can also directly be used in conjunction with foveation for further performance increase.

}

\CCScatlist{
  \CCScatTwelve{Computing methodologies}{Computer graphics}{Image compression}{};
  \CCScatTwelve{Computing methodologies}{Computer graphics}{Graphics systems and interfaces}{Virtual reality};
}

\usepackage[capitalize]{cleveref}
\crefname{section}{Sec.}{Secs.}
\Crefname{section}{Section}{Sections}
\Crefname{table}{Table}{Tables}
\crefname{table}{Tab.}{Tabs.}

\begin{document}


% \firstsection{Introduction}

\maketitle

\section{Introduction}

%% motivation / introduction codecs and video compression
Codecs provide efficient compression allowing to store hundreds of videos on a single drive.
This efficiency results from a precise adaptation to their specific application.
Video codecs like HEVC/H.265 or AV1 use the \ac{dct} and motion compensation for high compression rates at a reasonable perceptual quality.
The high compression, however, requires complex coding procedures.
The specific hardware decoders of modern graphics cards compensate for some of this decoding load.

%% solution: 360 videos
\vids are a sophisticated form of viewing experience which have become known with the spread of \ac{vr} technology.
In \vids the user can change their view anywhere, since the information of the entire space is available (three \ac{dof}).
This free exploration is not possible with traditional videos where the \ac{fov} is limited.
Accordingly, \vids are best suited for an immersive user experience for video playback in \ac{vr}.
%
%% VR video problem: huge amount of data
For comparable quality, the resolution of \vids must significantly exceed those of videos with a discrete view, since the representation on the display device only corresponds to a fraction of the whole video frame.
For a modern \ac{vr} headset with 2000x2000-pixel resolution per eye and 90\degree~\ac{fov}, a comparably resolved \vid would require 8000x8000 pixels (stereo) with up to 120Hz temporal resolution.
However, only the part of the frame that lies within the device's viewport at the time of decoding is relevant for rendering.
Current DCT-based codecs do not allow to load or decode only a defined part of a frame due to their complex non-linear structure.
Accordingly, with \vids the common practice is to load, upload, and decode the entire 360\degree~frame, even though only a small part of the frame is considered for the rendering. Furthermore, the recording quality of 360\degree~cameras is limited.
Consequently, the resolution and frame rate of \vids nowadays do not come close to the quality of the renderings of virtual environments.
%
%% problem RW
Although there are some existing ideas that already try to improve the compression -- e.g. tiling the frame into separate regions -- these approaches often go against the basic compression concept of \ac{dct} and are only a compromise at the expense of compression efficiency.

%% Wavelets
An alternative to the \ac{dct} for compression is the wavelet transform, which offers two decisive advantages over the former:
% In the 1990s, many researchers focused on the wavelet transform as a new compression option.\SUS{Cite}
% The wavelet transform offers two decisive advantages over the \ac{dct}:
(1) different parts of the image can be loaded and decoded individually, e.g. the viewport of an \ac{hmd};
(2) the encoding is performed in frequency layers, which are each halved in frequency. Decoding an area in fewer steps is equivalent to displaying the image area at a lower resolution.
Early attempts to use wavelet transform for image compression were limited to traditional presentations with a discrete \ac{fov} and have not gained wide use.

%% What we did
In this paper, we propose a wavelet-based codec for the compression of 360\degree~videos.
We particularly aim for high display speeds of high-resolution videos.
Our implementation of the wavelet-based codec uses the wavelet transform for inter- and intra-frame compression.
To the best of our knowledge, this is the first codec for \vids based on wavelet transforms.
In comparison with modern codecs (HEVC/H.265 and AV1) and related work, we show that our wavelet-based approach offers a significant speed advantage while providing a comparable video quality and reasonable compression rate.
In addition, we introduce foveated decoding. With foveated decoding the properties of the wavelets are used to gradually decrease the resolution with the distance from the focal point.
Such foveation is, so far, only known from virtual scenes and offers further opportunities to improve both decoding speed and image perception.
%
%\SUS{Can we actually share the codec in a compacted way or does it requiere Sascha's framework?}
Our codec will be made publicly available upon acceptance.

%\cg{This is also new:}
\textbf{The contributions} of the paper are summarized as follows:
\begin{itemize}
    \item a novel approach to encode and decode videos for fast viewport-dependent playback based on wavelet transforms
    \item wavelet-based inter-frame transform without keyframes  
    \item a technique to implicitly apply foveated rendering for wavelet-encoded videos during run-time
    \item objective and perceptual evaluation and comparisons with state-of-the-art video codecs and related work
\end{itemize}

\section{Related Work}

Loading the entire frame for \vid playback in VR is inefficient since only a fraction of the frame is actually rendered. However, this procedure is the most common practice, as it reduces the need to adjust the standard video pipeline.
In the following, we first discuss previous work aiming at a more resource-efficient presentation by adapting existing codecs and, in the second part, we introduce former attempts for wavelet-based codecs.

\subsection{Viewport-Adaptive Display Techniques for Videos}\label{sec:videoTechniques}

Zare\etal~\cite{Zare:2016} proposed to use a tiling scheme to increase the decoding speed of streamed 360\degree~HEVC videos.
Their experimental setup consisted of a pipeline with a dedicated server and client side. On the server side, the same video was encoded in high and low resolution. With the \ac{mcts} extension of HEVC the tiling was enabled for both versions of the video.
The client, on the other hand, requested the required tile sets from the server based on the viewport.
The authors tested different tiling schemes. The scheme with the most tiles ($18$ tiles) showed the highest bitrate savings ($-40\%$ based on Bjontegaard Delta Bitrate BD-BR~\cite{bjontegaard2001calculation}). However, the compression losses increase proportionally with the number of used tiles, since all tiles are saved independently.

The tiling approach was later officially formulated by the \ac{mpeg} into the \ac{omaf} standard for storage and distribution of 360\degree~videos~\cite{Choi:2018}. The idea of \ac{omaf} is comparable to the work of Zare\etal~\cite{Zare:2016} and is applied to HEVC or AVC video codecs.
The viewport-dependent streaming also uses \ac{mcts} to split the frames into tiles, each encoded in different qualities~\cite{Hannuksela:2019}.  

Sreedhar\etal~\cite{Sreedhar:2016} also recognized the technical challenges of bandwidth associated with high-resolution 360\degree~videos. The main focus of their work was the mapping techniques in which the recorded spherical scenes are packed in a rectangular frame. The most used mapping techniques are equirectangle and cubemap projection, which were also found as the most effective in their scenario. For the comparison, the authors presented a methodology of the rate-distortion performance of the schemes.

In the work of Corbillon\etal~\cite{Corbillon:2017} the 360\degree~videos are separated in individual tiles and offered in different resolutions. Unlike former works, the single tiles are created in different versions with only a selected part of every tile in a better visual quality.
While the 360\degree~video is streamed, the client communicates its viewpoint to the server which selects the tiles so that the viewpoint is in the higher quality region.
The paper does not specify actual display speeds, but it should be clear that the technology can save bandwidth.

%Further works propose saliency maps for an efficient decoding~\cite{nguyen2019saliency}.

%Afsana et al.
%Efficient Low Bit-Rate Intra-Frame Coding using Common Information for 360-degree Video

%Streaming methods:
%\citeauthor{Li:2021}~[\citeyear{Li:2021}]:
%A Log-Rectilinear Transformation for Foveated 360-degree Video Streaming

%%%%%%%%%%%%%%%%%%%%%%%%%%%%%%%%%%%%%%%%%%%%%%%%%%%
\subsection{Wavelet based codecs}
%%%%%%%%%%%%%%%%%%%%%%%%%%%%%%%%%%%%%%%%%%%%%%%%%%%

%\cg{fundamentals moved to supplemental; maybe short summary here}

Probably the best known use of wavelets for imagery is the JPEG2000 image compression standard~\cite{Taubman:2012, Marcellin:2000}.
At the turn of the millennium, it initiated a new form of image compression and was meant to replace DCT-based image compression formats.
%\SUS{Too strong formulation, rather leave out:}
%Unfortunately, the JPEG2000 format never reached the acceptance of the general public.
JPEG2000 supports lossless and lossy compression. The wavelet transform operates with the biorthogonal wavelets, either the \ac{cdf} $9/7$ wavelet~\cite{Cohen:1992} for lossy compression or the \ac{lgt} $5/3$ wavelet~\cite{LeGall:1988} for lossless compression.
The standard has four levels of decomposition as a default since there is not significant improvement in using higher decomposition levels when compressing images~\cite{JPEG2000:2019}.
One general advantage of wavelet compressed imagery is the progressive decoding, so that the quality of the visualisation improves progressively when more information is received. This progressive decoding is also supported in JPEG2000.

The JPEG2000 image standard was later extended to include video files. The extension is known as \textit{Motion JPEG2000} and is based on the MP4 format. This video standard uses the JPEG2000 coding for the compression of the individual frames. An inter-frame compression does not take place. Thus, Motion JPEG2000 is more of a container format for the joint wrapping of JPEG2000 compressed frames.

% Dirac %
Efforts to create video codecs based on wavelet compression are rare and nowadays exclusively experimental.
The most extensive attempt to create a wavelet-based video format to date was undertaken by BBC Research in 2008~\cite{DiracSpec}. The resulting versions of the codec were named \textit{Dirac} and \textit{Schrödinger} in honour of the Nobel Prize-winning physicists.
Dirac supports lossy and lossless coding for which it uses the same wavelets as JPEG2000 (CDF 9/7 wavelet or LGT 5/3 wavelet).
The motion compensation is performed with the \ac{obmc} logic for an effective inter-frame prediction~\cite{Orchard:1994}.
Unfortunately, this overlap also prevents effective intra-prediction, since there are no unique separations for overlapping blocks, as is the case with common DCT-based codecs.

However, the codec could not gain wide popularity and further development was discontinued more than a decade ago. The reasons for the codecs limited success are not entirely clear, but may be related to an inability to provide significant improvement over established codecs like H264.  
Dirac and Schrödinger are now abandoned and no longer available.

\begin{figure*}[th]
    \centering
    \includegraphics[width=\textwidth]{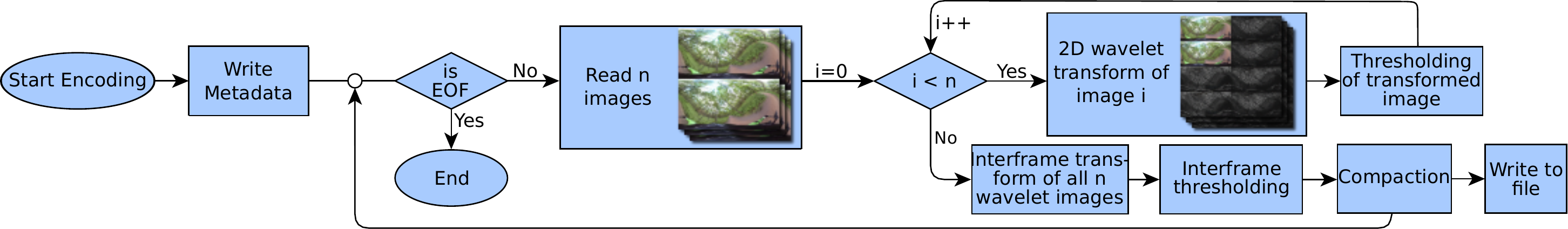}\\    \vspace{10pt}
    \includegraphics[width=\textwidth]{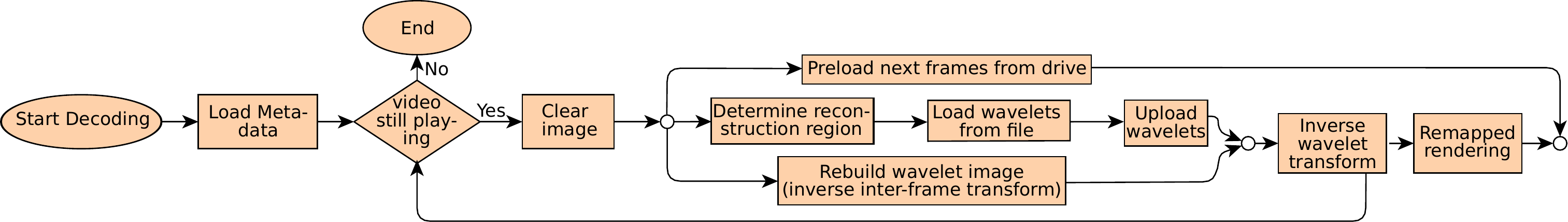}
    \caption{Our program flow for encoding (blue) and decoding (orange) a video with our wavelet-based codec.}
    \label{fig:programmFlow}
\end{figure*}

\section{Method}

%Video coding is highly complex.
Two concepts that most video codecs apply for data compression are intra- and inter-frame coding.
In practice, these methods are commonly applied with some information loss to achieve better compression ratios.
%\frg{maybe cite work that shows that lossy compression achieves better compression ratios.}
\textit{Intra-frame coding} usually refers to the transformation of the data of one frame to a different representation that can be compressed more efficiently.
\textit{Inter-frame coding} utilizes redundancies between multiple frames to reduce the data size. %In modern video codecs this is done with the use of key frames.
In the following, we describe how we realised both concepts with wavelet transforms.

% \begin{figure*}[th]
%     \centering
%     \includegraphics[width=.48\textwidth]{images/Flowchart Encode V3.pdf}
%     \includegraphics[width=.48\textwidth]{images/Flowchart Decode V2.pdf}
%     \caption{Our program flow for encoding (blue) and decoding (orange) a video with wavelet-transforms.}
%     \label{fig:programmFlow}
% \end{figure*}

%%%%%%%%%%%%%%%%%%%%%%%%%%%%%%%%%%%%%%%%%%%%%%%%%%%%%%%%%
\subsection{Frame-wise Transform}\label{sec:transform}
%%%%%%%%%%%%%%%%%%%%%%%%%%%%%%%%%%%%%%%%%%%%%%%%%%%%%%%%%

The core of the frame-based compression of our codec is a 2D \ac{fwt}.
Similar to other codecs, the transform of the frame data allows for a better compression, which in the raw state is too large to be stored.
For example, a one minute \vid in 8k resolution would contain around $300 \text{GB}$ uncompressed data.  
We transform an input frame $s$ of $(N\times M)$ pixels for a discrete 2D position $(x,y)$ and frequency $\gamma$ by the wavelet transform $W$ with the mother wavelet~$\psi$.
\begin{equation}
    W_\psi s(\gamma,x,y) = \sum_{i=0}^N\sum_{j=0}^M \psi_{\gamma,x,y}(i,j) s(i,j)
\end{equation}
%
%Generally, an image that is transformed by a wavelet transform has the same number of wavelet coefficients as it has pixels.
To compress the transformed frame $W_\psi s$ all coefficients below a certain threshold $T_s$ are set to zero, so that:
\begin{equation}
    W'_\psi s \coloneqq
     \begin{cases}
        W_\psi s, & if \lvert W_\psi s \rvert  > T_s, \\
        0, & \text{otherwise} \\
     \end{cases}
\end{equation}
For a properly chosen $T_s$, this operation has only minor implications for the quality of the reconstructed image.
This is especially true for high frequencies and is a general characteristic of the frequency domain.
Usually, in natural images most of the information is contained in the low frequencies, which are represented by only a small number of coefficients~\cite{Unser:2003}.
%By utilising this advantage of the frequency domain, the wavelet transform becomes an effective compression method we can use for video encoding.
As usual for the \ac{fwt}, with every step of the 2D transformation, the resolution of the image approximation is halved in both dimensions. In $W_\psi$ this is addressed by the frequency layers over $\gamma$.
% \SUS {This should be the core of the technical part, needs proper description}

Former research has shown that the discrete wavelet transform can achieve better image reconstruction than a \ac{dct}-based method at high bit compression ratios~\cite{Boopathi:2012}.
For the frame-wise transformation of the image we use the \ac{cdf} 9/7 wavelet~\cite{Cohen:1992}.
The \ac{cdf} wavelet is known to perform especially good on natural imagery and is also used for lossy compression in the JPEG2000 standard~\cite{JPEG2000:2019}. %The ability of the CDF to produce high-quality images with only few coefficients is further demonstrated in our own experiments.

During playback, the compressed video information is decoded with an \ac{ifwt} to obtain the original images.
This reconstruction is not conducted for the entire image, but only for the part of the 360\degree~panorama that lies in the viewport of the display device.
For a viewport-dependent reconstruction, we define the location of the viewport on a low resolution representation of the frame in binary form. This binary mask is uploaded together with the wavelet coefficients and is used for the inverse wavelet transform. The resolution of the binary representation is $256\times256$ pixels for a 8k stereo frame and can define arbitrary reconstruction shapes.
%\frg{and}
%The part of the frame that is to be reconstructed is defined by a low resolution binary mask.
%\SUS{Which binary mask}
% This binary mask is defined by the wavelet coefficients that are loaded from the drive.
% From the reconstruction it is necessary that the reconstructed areas do not become larger on higher levels. Otherwise the low-frequency information would be partly missing and the result would not be visually plausible. In our case this constraint is always kept.
%The inverse wavelet transform is only performed for the masked area and, therefore, viewport-dependent.

%% Lowest Frequency
In theory, the transform can be performed until only one pixel defines the lowest frequency over the whole image.
However, the low-frequency levels of the wavelet transform contain fewer discrete data points since the high-resolution in the frequency domain results in a low resolution in time due to the Heisenberg theorem~\cite{Heisenberg:1927}. Also, the wavelet coefficient values of these pixels can only be compressed inefficiently because the low-frequency information is significantly more important than high frequencies in natural image reconstruction.
By default, we perform the wavelet transform for $l_{max} = log_2(\frac{N}{32})-2$ levels.
For the number of wavelet levels, we took inspiration from the JPEG2000 standard, but also performed several pilot studies. While we found $6$ levels to be the optimal default solution for 8k videos, the number of wavelet levels can be chosen individually per video.

%%%%%%%%%%%%%%%%%%%%%%%%%%%%%%%%%%%%%%%%%%%%%%%%%%%%%%%%%
\subsection{Inter-Frame Coding}\label{sec:interframe}
%%%%%%%%%%%%%%%%%%%%%%%%%%%%%%%%%%%%%%%%%%%%%%%%%%%%%%%%%

Inter-frame coding describes the compression of temporal information. In videos, the time component $t$ is represented implicitly by a set of successive frames.
In modern codecs the inter-frame compression is performed with keyframes and motion vectors where only the information differences are encoded. While this technique offers impressive compression rates, a compression with keyframes has the disadvantage that its speed depends on the linear information retrieval. When the video is skipped, all information since the last keyframe has to be reloaded first.
With our codec we wanted to get rid of this disadvantage and at the same time maintain a good compression rate between inter-frames. To achieve this purpose we apply a second one-dimensional wavelet transform that encodes the temporal pixel differences.
The second wavelet transform is applied on a set of wavelet images resulting from the frame-wise wavelet transform $W_\psi s$. Here, we use a one-dimensional form of $W_\psi$ with $s(\gamma, t)$ for the frequency $\gamma$ of the temporal changes of every pixel over time. In other words, our wavelet based inter-frame transform encodes the frequency changes over time.
Typically, even with movement in the frame the temporal information only changes on single frequencies. All frequencies that do not or only slightly change are compressed by our inter-frame transform.
As result, the speed of the decoding is unaffected by the direction in which the viewport moves.
In theory both, the frame-wise transform and the inter-frame transform, can be combined to one 3D wavelet transform. However, this 3D transform would not offer us the possibility to decode different areas of a frame in different resolutions for the same computational costs. Furthermore, the separation allows us to apply different wavelets and thresholds per transform and respond adaptively to individual circumstances.

Every inter-frame transform of \textit{n} consecutive frames we call \textit{inter-frame set}. Thereby, n is a power of two value. The number of frames per inter-frame set can be defined per video and may be bigger the less motion is in the video.
In contrast to the frame-wise transform, the inter-frame transform is always executed to the last level.
For the inter-frame wavelet transform we use the Haar wavelet~\cite{Haar:1911}. The Haar wavelet is the only wavelet with no overlapping of the wavelet filters and can therefore be reconstructed by loading only one coefficient per level for the high and low pass filtering. Reconstruction of one specific pixel by a Haar wavelet transform with $n$ levels only requires $log_2(n)$ additions of the correct wavelet coefficients multiplied by the high pass filter position (either $-1$ or $1$).
As a result, for the inverse inter-frame transform we can iterate over the uploaded wavelets rather than over all pixels of the target section.
This characteristic is unique to the Haar wavelet and allows a rapid inter-frame reconstruction.
The speed of the inter-frame reconstruction is important since the inverse inter-frame transform runs on $log_2(n)$ frames every time one frame is decomposed.
By iterating over the uploaded pixels rather than a target section we implicitly synthesise only the part of the image that is in our defined \ac{fov}. 
%This area is equal to the one denoted by the binary reconstruction mask used for the frame-wise reconstruction (cf. Sec.~\ref{sec:transform}).
%With our codec, the inverse inter-frame transform takes up about X percent of the reconstruction time.

%%%%%%%%%%%%%%%%%%%%%%%%%%%%%%

\begin{figure*}
    \centering
    \includegraphics[width=.8\linewidth]{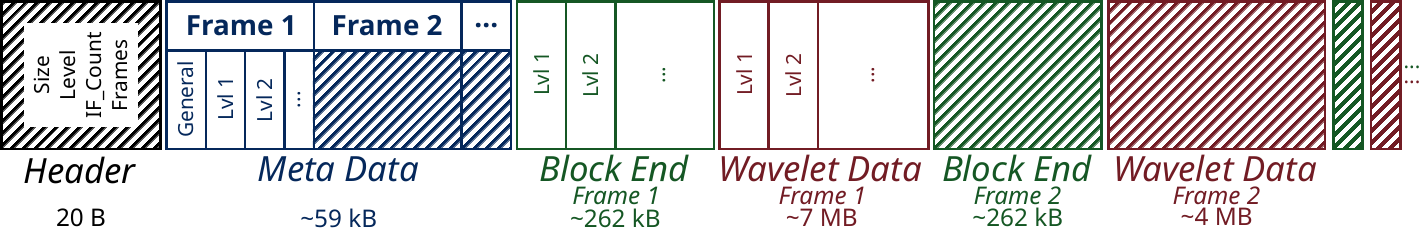}
    \caption{Data arrangement of our video format. The sizes of each section are given by an example video in 8k resolution.}
    \label{fig:fileFormat}
\end{figure*}

%%%%%%%%%%%%%%%%%%%%%%%%%%%%%%%%%%%%%%%%%%%%%%%%%%%%%%%%%
\subsection{Thresholding}
%%%%%%%%%%%%%%%%%%%%%%%%%%%%%%%%%%%%%%%%%%%%%%%%%%%%%%%%%

In order to achieve the necessary storage savings, we have to determine which wavelet coefficients are least essential for the reconstruction. We will refer to this step as \textit{thresholding}. The chosen threshold value is decisive for the intensity of the compression. Thereby, the threshold always represents a trade-off between quality of the reconstructed image and size of the video file. The threshold of 0 represents lossless compression while a threshold of 1 produces the smallest file. Reconstructing an image with too little frequency information may result in a blurry representation with less details.
We derive a threshold $T$ from a user-defined constant $\alpha$ and the level $l$ of the transform:  %$H$ is further defined in Section~\ref{sec:mapping}.
\begin{equation}\label{equ:thres}
    T(x,y) = \alpha \biggl(\frac{l_{max} - l}{l_{max}}\biggr)^2 + H,
\end{equation}
where $H$ specifies a mapping factor which depends on the mapping technique (cf. Sec.~\ref{sec:mapping}).
%In the thresholding process, all coefficients that are closer to zero than the chosen threshold $T$ are zeroed out and are not saved in the video file.
In the encoding, the frames are thresholded twice: once after the frame-wise 2D \ac{fwt}, and again following the inter-frame transform.
We use two separate threshold operations as both wavelet transforms aim for a different encoding: The frame-wise transform encodes the different frequency information in the respective spatial resolutions. The inter-frame transform encodes temporal frequency information of every wavelet coefficient. %pixel.
%\frg{temporal/frequency information of inter-frame wavelet coefficients, not pixels.}
Thresholding the values only once is possible but, in our experience, can lead to unwanted interactions and a worse compression rate.
%\frg{What is this supposed to mean ?!}
%
Both thresholding operations are independent and have their own %shader and
threshold value. While the first thresholding is applied on every frame independently, the thresholding of the inter-frames considers all $n$ frames of the inter-frame set. In this latter thresholding operation, the different levels are defined by the relative frame number from $t$ rather than the pixel position inside the frame. % (cf. Sec.~\ref{sec:interframe}).
%\frg{i would not talk about ''shaders'' or ''compute threads'' etc. you never established any implementation environment prior to this and also this chapter is about the method, not the implementation details, frameworks and libraries used.}

High frequency information was found to be less important for the perceptual quality of an image than low frequency information~\cite{Unser:2003}.
We scale the threshold by the frequency level of each coefficient in a quadratic function (cf.~Eq.~\ref{equ:thres}). Accordingly, more coefficients may be zeroed out at high frequencies.  
This thresholding weighting follows common procedures of other codecs like JPEG2000~\cite{Taubman:2012, Marcellin:2000}.

\textbf{Quantization:}
Similar to other codecs, we represent the color values of the pixels in the video file by one byte per color component.
In the quantization, the $32$ bit float color components of the wavelet transform are mapped to the byte representation of the compressed output. We use the extreme values of the wavelet coefficients for normalization in order to achieve the highest possible spatial resolution in this discretization.
Therefore, one discrete color value $c_d$ is defined by 
\begin{equation}
    c_d = \frac{c_n - c_{min}}{c_{max} - c_{min}} * 255    
\end{equation}
 
with $c_n$ as the floating-point representation of the $n$-th pixel and $c_{min}$ and $c_{max}$ as the minimum and maximum values of all coefficients, respectively.
The normalization is performed with a separate minimum and maximum for the approximation area (last layer of the transform) and the wavelet layer and for each inter-frame.
The normalization is inverted during reconstruction. The minimum and maximum values are stored with the metadata in the file.

%%%%%%%%%%%%%%%%%%%%%%%%%%%%%%%%%%%%%%%%%%%%%%%%%%%%%%%%%
\subsection{File Format}
%%%%%%%%%%%%%%%%%%%%%%%%%%%%%%%%%%%%%%%%%%%%%%%%%%%%%%%%%

The structure of our video file is illustrated in Fig.~\ref{fig:fileFormat}.
We designed the layout to allow for a fast and viewport-dependent streaming of the data.
Starting with a file header, general information on the video is offered. This data includes the number of frames, size of the frames and number of levels of the wavelet transform.
Following the header, metadata information on every single frame is provided.
This frame-wise metadata includes information about where the frame starts and ends in the file or the overall number of wavelets. The frame metadata also provides information on the individual levels of this frame.
The header as well as the entire metadata are preloaded when the video is started and are kept in the working memory.

%block segmentation
%For our video codec, the wavelet transform allows us to access arbitrary areas of a frame, e.g., to reconstruct only the viewport of a VR user.
The position $(x,y)$ of one particular wavelet coefficient within the frame is given by an index that is saved along with the wavelet value.
However, due to the compression, the position of individual coefficients within the file is unknown.
While it is possible to find the data for $(x,y)$ with binary search on the video data, this inconsistent access to the storage drive adds an unwanted delay to the loading process.
Instead we divide the transformed frames into a logical grid of small blocks (default size $32\times32$-pixel).
This block allocation is only relevant for the compression but does not affect the wavelet transforms which operate on the entire images. Note, that this is different from blocking in \ac{dct}.
In the video file we store one pointer for each block, located in the \textit{BlockEnd} section (see Fig.~\ref{fig:fileFormat}). This pointer indicates where the last wavelet coefficient inside the respective block can be found in the video file.
In the file the coefficients are stored block after block, which allows a whole series of blocks to be loaded by two of these block-end pointers.
During decoding, the block pointers of a frame are preloaded before the frame is processed. % (more in Sec.~\ref{sec:processes}).
%
% Potentially, a slight exceeding can result from this block-wise loading, when a block lies only partly in the reconstructed part.
% However, such cases are not handled in a special way, as the upload of more wavelets affect the speed of the decoding less than a dedicated processing of such cases, given a reasonable block-size.
%
%The two sets of the block-ends and wavelet coefficients of one frames are written one after the other to the video file (cf. Fig.~\ref{fig:fileFormat}).
%For the encoding, not all frames are loaded at the same time since this could cause the RAM to overflow.
By alternately storing the block-end and wavelet data packages of the frames in the video file we can avoid compute-intensive rearrangements of the file during encoding.
Inside one block of wavelet coefficients or block-end information the data is ordered level-wise starting with the lowest frequency layer.

%%%%%%%%%%%%%%%%%%%%%%%%%%%%%%%%%%%%%%%%%%%%%%%%%%%%%%%%%
\subsection{Parallel Wavelet Processing}\label{sec:processes}
%%%%%%%%%%%%%%%%%%%%%%%%%%%%%%%%%%%%%%%%%%%%%%%%%%%%%%%%%

%The part of surrounding material that is displayed on the viewport, e.g. a VR display, is only a fraction of the full \vid frame.
% This segmentation allows the user to look around in the scene and have a higher immersion. It also means for a surrounding frame that a lot of data is uploaded and decoded unnecessary, when the whole frame is considered.
% Wavelet based coding methods allow for a simpler selection of the required data for uploading and decoding since the encoding acts pixel-wise.
%

Since VR users move their head, the part of a \vid that is rendered can change continuously. These viewpoint changes complicate the buffering of subsequent frames with a viewport-dependent video stream. 
However, buffering is necessary to avoid load peaks and latencies that disrupt the virtual experience and can induce cybersickness~\cite{Stauffert:2020}. 
%On the other hand, without buffering the data has to be loaded from the drive at the time of rendering which is costly and slows down the process. 
%
Therefore, we load the viewport data of the next frames asynchronously while the current frame is decoded. 
Until the time of rendering, all frames are updated continuously in case that the look direction changes.
%
%\cg{\textit{Maybe include:} Given the high refresh rate of the HMDs and the maximum movement speed of the human head, we are able to avoid any additional latency due to our asynchronous loading process.}
%
The number of frames that are preloaded corresponds to the inter-frame size. Preparing more frames is not always useful, as the view direction may change strongly over longer periods of time.
During our experiments no substantial latency was measured that arose from the buffering.
%
%At the time of rendering, in the case that some data of the current frame is missing, the preloading process is stopped until the loading is completed (see \textit{wait} sections in Fig.~\ref{fig:processes}).

\begin{figure}
    \centering
    \includegraphics[width=\columnwidth]{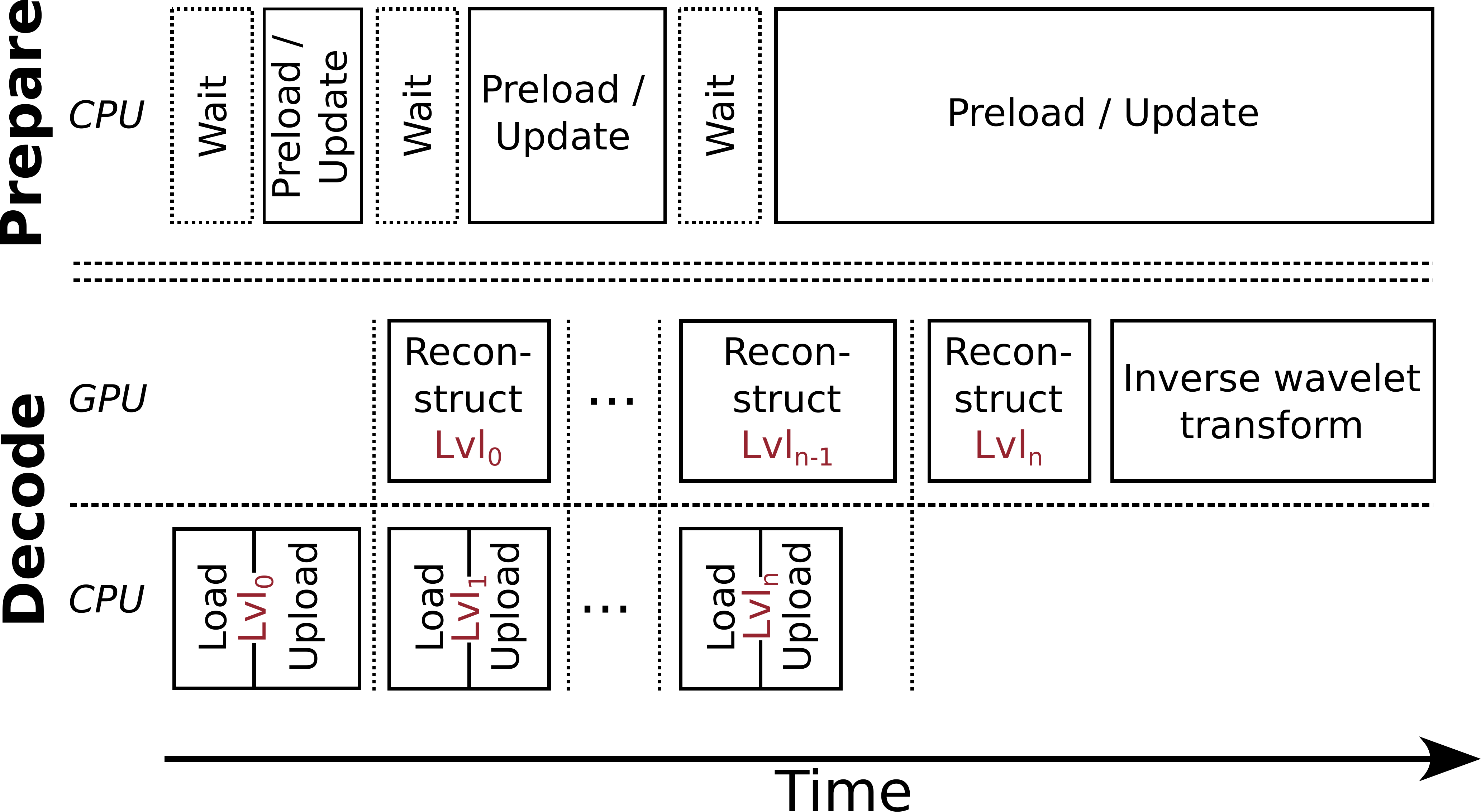}
    \caption{Parallel processes for the reconstruction of one frame. The prepare thread buffers the next frames. In the \textit{wait} sections the drive is occupied by the decoding process.}
    \label{fig:processes}
\end{figure}

Due to the inter-frame compression, one frame is reconstructed by the wavelet coefficients stored in multiple inter-frames of the respective inter-frame set.
%Consequentially, for the reconstruction of one part of a frame, the same part has to be loaded from all relevant inter-frames.
%The number of relevant inter-frames is $log_2(n)$ for a set of $n$ inter-frames (more details in Sec.~\ref{sec:interframe}). However, this exceeding is compensated by the larger compression rate.
We rebuild the wavelet representation $W_\psi s$ of one frame on the GPU while we already upload the wavelet data for the next inter-frame in parallel (see Fig.~\ref{fig:processes}).  
This rebuild already includes the inverse inter-frame transform, as described in Section~\ref{sec:interframe}.
% Also, the frame-wise \ac{ifwt} is executed on the GPU.
The reconstruction of the original frame by the 2D \ac{ifwt} is performed once all inter-frames are processed and the inverse inter-frame transform is completed.

%%%%%%%%%%%%%%%%%%%%%%%%%%%%%%%%%%%%%%%%%%%%%%%%%%%%%%%%%
\subsection{Frame Mapping}\label{sec:mapping}
%%%%%%%%%%%%%%%%%%%%%%%%%%%%%%%%%%%%%%%%%%%%%%%%%%%%%%%%%

\vids are representations of a recorded 3D sphere, brought to a rectangular frame by a projection. The position of the information in the projected frame is defined by its mapping.
Among the most popular mapping techniques are equirectangular mapping and cubemaps.
%
%Equirectangular images are easy to display and human-interpretable, like conventional images. The problem of equirectangular projections is their information density, which is not uniformly distributed over the image, but increases from the equator to the poles. Thus, the entire topmost pixel row of a mapped frame represents only one point in the 3D scene.
%Cubemaps, on the other hand, are not so easy to observe in their projected form. Instead they provide a better distribution of the information density. The distribution of information is particularly uniform for Equi-Angular Cubemaps~\cite{GoogleECA}.
%
Our codec is defined to be independent of the mapping technique. As the reconstructed areas are given in a low resolution presentation of the frame (cf.~Sec.~\ref{sec:transform}), the areas needed for the frame mapping can be set in direct relation to the final reconstruction.
%The mask defines which pixels have to be reconstructed and is directly dependent on the wavelets that are loaded.  
We render the final re-projection of the \ac{fov} to the spherical presentation in an own shader which is run after the inverse wavelet transform is completed. This shader can react on multiple mapping types and also considers the stereo images.
For the experiments we use the equirectangular projection.
We tackle the redundancies in the pixel information near the poles by gradually increasing the mapping factor $H$ towards the poles.
For an equirectangular projected frame with the dimension $S$ we define $H(y) = 1 - \sin(y \pi / S_y)$.
%This factor is calculated by the cosine of the squared lateral distance from the equator.
With the adjusted threshold, we experience equal performance at all viewing angles, including upward views.

%%%%%%%%%%%%%%%%%%%%%%%%%%%%%%%%%%%%%%%%%%%%%%%%%%%%%%%%%
\subsection{Foveated Decoding}\label{sec:foveation}
%%%%%%%%%%%%%%%%%%%%%%%%%%%%%%%%%%%%%%%%%%%%%%%%%%%%%%%%%

\begin{figure}
    \centering
    \includegraphics[width=\columnwidth]{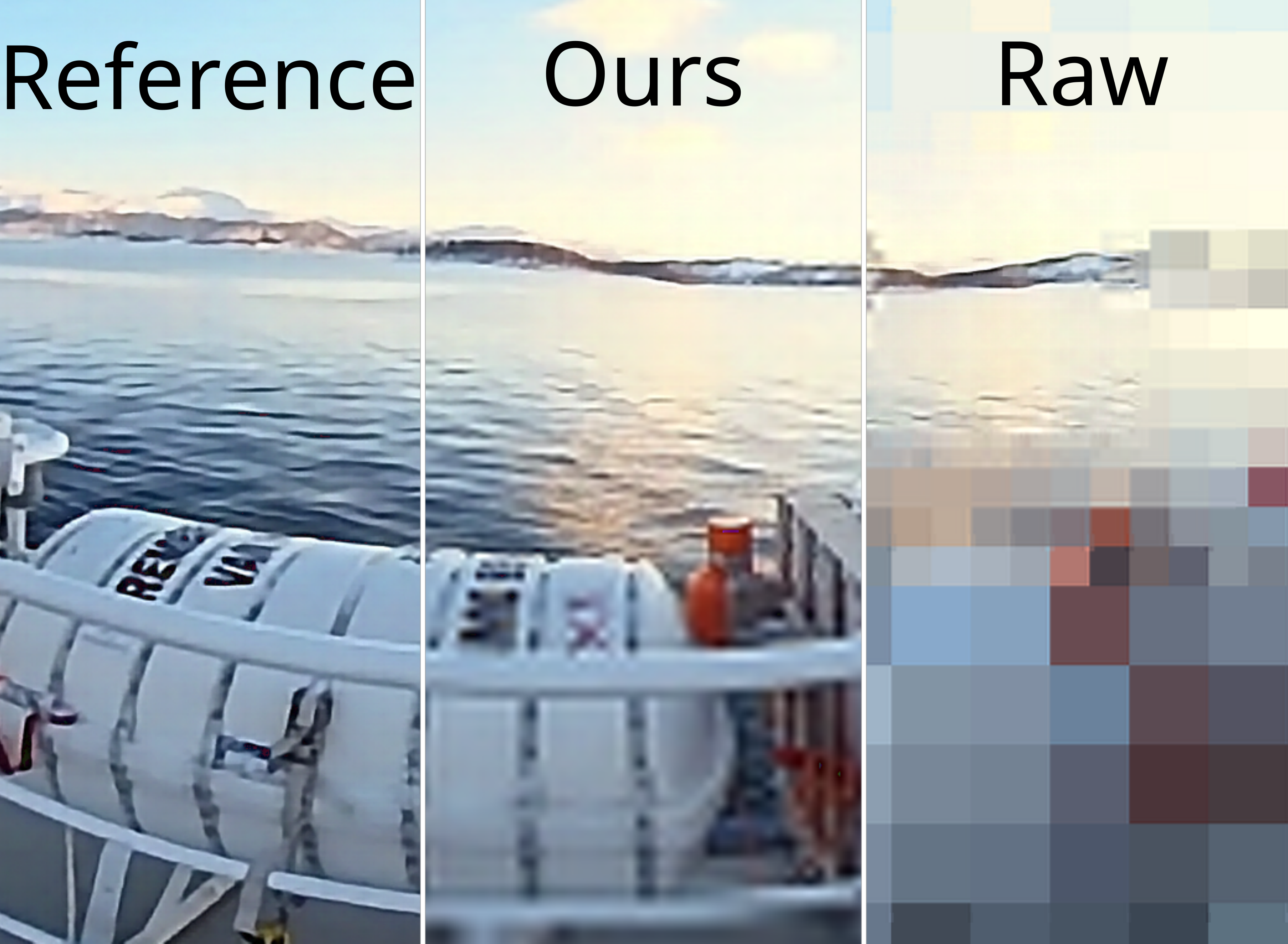}
    \caption{Our foveated decoding compared with the full-resolution reference frame. On the right, the information density of the foveated decoding is visualised.}
    \label{fig:foveationExample}
\end{figure}

The information density of visual representations of the human eye are not equally distributed~\cite{Silverstein:2008}. The images created on the retina of the human visual system follow a qualitative decline starting from the eye's fixation point. %which are further processed in the visual cortex
While people can perceive the full resolution of about one sixtieth of a degree in the fovea around the focus point, the information in the peripheral visual area is significantly lower in resolution~\cite{Kolb:2020}.

So far we only discussed full resolution reconstructions of the viewport.
However, when the eye gaze direction of the observer is available by eye tracking, we can utilise the properties of the wavelet transform to achieve what we call \textit{foveated decoding}.
With foveated decoding the resolution gradually decreases with the distance from the fovea. Our method is comparable to classical foveated rendering, except that in the periphery the decoding is accelerated while the rendering load is constant.
The results are bandwidth savings and higher playback speeds.

For the foveated decoding, we utilise the level-wise structure of the wavelet transform. As described in Section~\ref{sec:transform}, a wavelet representation is composed of individual levels, each corresponding to a defined frequency interval $\gamma$. The reconstruction is performed incrementally from a low-resolution version of the frame to the full resolution.
Instead of reconstructing the same \ac{fov} at each level, the full \ac{fov} of the viewport is reconstructed only at the lowest frequency level. After that the \ac{fov} is reduced with every level resulting in a compact foveal area.
The sizes of the individual resolution levels of the wavelet transform are defined in regard to the properties of the human eye~\cite{Leigh:2015}. Like human perception, we decrease the quality of the frames at a logarithmic rate~\cite{Kolb:2020}.
The area with the full video resolution which stimulates the most central foveola is only about two percent wide~\cite{Silverstein:2008}.

The inverse wavelet transform is executed over the same number of data points as for a full resolution viewport but assumes the coefficients to be zero for the surrounding regions. With foveated decoding, we achieve a peripheral reconstruction in a visually appealing quality with a small number of coefficients (see Fig.~\ref{fig:foveationExample}).
%The uploaded data corresponds to the decreasing \ac{fov}.
With the foveation, up to 80\% less data has to be loaded.  
The reconstructed areas are in rectilinear form and follow recent findings, which indicate advantages over a log-polar presentation~\cite{Li:2021}.

%\cg{TODO Predictions of eye gaze trajectories}
%\cg{TODO detect saccades and transfer less data}

\section{Experiments}

% Unfortunately, former wavelet-based video codecs like Dirac were not available anymore and therefore cannot be included in the evaluation.
In objective and subjective evaluations we compare our codec against the common HEVC and AV1 codecs. The benchmarks for display speeds also include a tiled HEVC implementation from related work~\cite{Zare:2016}.  

\subsection{Dataset}\label{sec:stimuli}

For the evaluation, we aim to analyse a particularly diverse set of videos.
Both \ac{cgi} content and real world recordings are considered. Furthermore, we investigate moving camera trajectories as well as fixed recording positions. 

The \ac{cgi} reference frames are created with \textit{Unreal Engine 5} with a high level of photorealism. 
The \ac{cgi} scenes cover urban and nature scenery. The \textit{City} scene mainly contains geometrical structures and straight lines and the recording takes place in a virtual New York City.
The \textit{Mountain} scene depicts a natural landscape with trees and a lake. The content of this scene contains a more heterogeneous shape composition compared to urban imagery.
In the \ac{cgi} scenes the camera trajectory moves along a predefined quarter circle path. The frames are rendered as 360\degree~stereo images with $8192\times8192$ pixel resolution.

%For the evaluation, we analyse two categories of 360\degree~videos.
The first set of real-world videos is recorded with a moving camera trajectory and the display of rapid motions. Here, we use the videos of Groth and colleagues~\cite{Groth:2022, Groth:2021} which have a higher resolution than typical moving-camera \vids due to their custom camera setup.
The second category considers videos with a fixed recording position (further denoted by~*). These videos were originally recorded by M\"{u}hlhausen\etal~\cite{Muehlhausen:2020}.
%All videos display natural, real-world scenes. We deliberately decided not to use rendered scenes, as we see the final application area to be real-world recordings. 
The original real-world videos are recorded with stereoscopic information in $6400\times6400$-pixel resolution at 30 \ac{fps}.

\textbf{Reference Data Creation:}
Pre-recorded videos can cause two problems for the evaluation. For one, the frame rates typically do not match the refresh rates of VR devices. Additionally, the data is already lossy compressed and can include serious compression artifacts.
In order to address both problems, we first downscale the video data to $1024\times1024$-pixel to get rid of high-frequency compression artifacts and then perform temporal interpolation and upscaling of the data with \acl{sota} neural network approaches.
The resulting frames are used as reference for our evaluation.
The original video data is downscaled with bicubic interpolation by \textit{OpenCV}.
%This interpolation method produced the best results for the data from visual observation.
The temporal interpolation is run on the downscaled frames with RIFE~\cite{Huang:2021}. The information from both eyes is processed individually to avoid artifacts at the edge. We increase the frame rate from the original 30 \ac{fps} to 120 \ac{fps}, which should be in line with the frequency of most modern VR glasses.
For the resolution upscaling we use Nvidia VFX to create the final reference frames in $8196\times8196$-pixel resolution.

%%%%%%%%%%%%%%%%%%%%%%%%%%%%%%%%%%%%%%%%%%%%%%%%%%%%%%%%%%%%%%
\subsection{Objective Evaluation}
%%%%%%%%%%%%%%%%%%%%%%%%%%%%%%%%%%%%%%%%%%%%%%%%%%%%%%%%%%%%%%

\begin{table}
\centering
\caption{\textbf{Display speeds.} Tiling refers to Zare\etal\cite{Zare:2016}. The foveated decoding (\textit{FD}) is run with the high resolution version of our codec ($Ours_{HQ}$), all values are averages over multiple runs and given in \ac{fps}.}
\label{tab:speed}
\centering
\setlength\tabcolsep{6.5 pt}%9
{\small%\footnotesize%\scriptsize%
\begin{tabular}{p{1.1cm}llllll}
\toprule
\multirow{2}{\linewidth}{Videos}    & \multicolumn{6}{c}{AVG fps $\uparrow$} \\
                 & HEVC    & Tiling & AV1    & $Ours_{LQ}$ & $Ours_{HQ}$& $Ours_{FD}$  \\
\midrule    
City$_{CGI}$     & 57.32   & 88.24  & 52.7   &  220.96    &   194.17    &  \textbf{228.8}      \\
Mountain$_{CGI}$ & 58.27   & 87.95  & 61.57  &  222.75    &   202.28    &  \textbf{227.46}      \\
Downhill         & 60.21   & 93.21  & 48.32  &  193.88    &   180.70    &  \textbf{206.65}     \\
Horse            & 62.6    & 95.94  & 50.45  &  197.17    &   181.18    &  \textbf{207.16}      \\
Climbing         & 65.36   & 92.55  & 52.53  &  195.63    &   187.56    &  \textbf{207.3 }     \\
Walking          & 65.32   & 93.1   & 52.76  &  198.24    &   187.88    &  \textbf{205.89}      \\
Cave*            & 66.53   & 111.08 & 53.93  &  209.22    &   207.12    &  \textbf{210.28}      \\
Boat*            & 67.13   & 110.15 & 55.07  &  205.21    &   201.06    &  \textbf{208.76}      \\
\bottomrule          
\end{tabular} 
}
\end{table}

We consider a high and low quality version of the wavelet-compressed videos.
The inter-frame transform is applied in sets of four frames and compressed with an inter-frame threshold of $0.005$. The frame-wise threshold is chosen to be $0.1$ and $0.25$ for the high and low quality version, respectively.
The HEVC and AV1 encodings are performed with FFmpeg.
Regarding quality, for HEVC we use a \ac{crf} of 30 (range 0--51) and for AV1 a \ac{crf} of 50 (range 0--65). The videos of both codecs are encoded in the common YUV420 colour space.
The \ac{omaf} inspired tiling method is realised with HEVC encoded tiles with the fastest tiling scheme of Zare\etal~\cite{Zare:2016}. However, we extended their tiling scheme for stereoscopic videos to a 6-by-6 grid layout (6-by-3 per eye). Following the former work, the middle row of both eyes is chosen with 90\degree~height and all other rows with 45\degree~height for a better central view performance.

We conduct all of our experiments on a commercially available computer with a NVIDIA RTX 3090 graphics card and an AMD Ryzen 5950X processor. The video data is stored on an on-board SSD. A HTC Vive Pro Eye is chosen as output device.
All videos are displayed in our self-programmed video player which uses the Vulkan API to utilise the GPU.
The decoding of HEVC and AV1 video data is performed with the Nvidia NVDECODE API. Thereby, the HEVC and AV1 decoding benefits from the hardware acceleration on the GPU. All videos are created from 1200 reference frames with 8192x8192-pixel resolution.
To assure an equal comparison of all experimental conditions, we use head and eye tracking data of participant recordings.

%\cg{height factor not used for quality analysis - would falsify the results}

% \begin{figure}
%     \centering
%     \includegraphics[width=\columnwidth]{images/QualityComparison.pdf}
%     \caption{Visual comparison of the codecs by different sections of the image.}
%     \label{fig:artifactCompare}
% \end{figure}

%\input{tables/tables}

\begin{table}
\centering
\caption{\textbf{Objective Quality Comparisons}. The results from the computational metrics, PSNR/SSIM (higher is better) and LPIPS (lower is better), on all codecs compared with the reference frames.}
\label{tab:quality}
\centering
\setlength\tabcolsep{9 pt}%9
{\small%\footnotesize%\scriptsize%
\begin{tabular}{ccccccc}
\toprule
Scene& Metrics       & HEVC             & AV1        & $Ours_{LQ}$ & $Ours_{HQ}$  \\
\midrule
\multirow{3}{1cm}{City$_{CGI}$}
&PSNR $\uparrow$     &    35.53          &  33.32    &  29.46        & 32.51  \\
&SSIM $\uparrow$     &    .905           &  .89      &  .817         & \textbf{.916}   \\
&LPIPS $\downarrow$  &    .124           &  .167     &  .268         & \textbf{.114}   \\
\midrule  
\multirow{3}{1cm}{Mountain$_{CGI}$}
&PSNR $\uparrow$     &    36.08          &  35.69    &  30.57        & 33.09  \\
&SSIM $\uparrow$     &    .927           &  .925     &  .851         & \textbf{.933}   \\
&LPIPS $\downarrow$  &    .158           &  .157     &  .32          & \textbf{.145}   \\
\midrule  
\multirow{3}{1cm}{Downhill}
&PSNR $\uparrow$     &    34.77          &  34.17    &  31.4         & \textbf{34.81}  \\
&SSIM $\uparrow$     &    .954           &  .95      &  .921         & \textbf{.961}   \\
&LPIPS $\downarrow$  &    .082           &  .103     &  .161         & \textbf{.081}   \\
\midrule               
\multirow{3}{1cm}{Horse}
&PSNR $\uparrow$     &  \textbf{37.05}    &  36.48    &  32.24     &  35.07           \\
&SSIM $\uparrow$     &  .968              &  .966     &  .939      &  \textbf{.97}           \\
&LPIPS $\downarrow$  &  \textbf{.054}     &  .07      &  .118      &  .057           \\
\midrule
\multirow{3}{1cm}{Climbing}
&PSNR $\uparrow$     &    \textbf{37.3}   &  36.83    &  32.84       &   35.33          \\
&SSIM $\uparrow$     &   .973             &  .971     &  .952        &  \textbf{.976}           \\
&LPIPS $\downarrow$  &   \textbf{.051}    &  .066     &  .115        &  .056           \\
\midrule
\multirow{3}{1cm}{Walking}
&PSNR $\uparrow$     &   \textbf{37.64}   & 37.06     &  32.04       &  35.45           \\
&SSIM $\uparrow$     &   \textbf{.97}     & .969      &  .928        &  .967           \\
&LPIPS $\downarrow$  &   \textbf{.05}    & .064       &  .117        &  \textbf{.05}           \\
\midrule
\multirow{3}{1cm}{Cave*}
&PSNR $\uparrow$     &  \textbf{40.32}    & 40.24          &  37.02    &  39.58           \\
&SSIM $\uparrow$     &  .971              & \textbf{.972}  &  .96      &  \textbf{.972}     \\
&LPIPS $\downarrow$  &  \textbf{.039}     & .04            &  .08      &  .046           \\
\midrule
\multirow{3}{1cm}{Boat*}
&PSNR $\uparrow$     &  39.2              &  \textbf{39.87}    &  33.01         &  35.88           \\
&SSIM $\uparrow$     &  .984              &  \textbf{.986}     &  .954          &  .978          \\
&LPIPS $\downarrow$  &  .03               &  \textbf{.027}     &  .078          &  .036           \\
\bottomrule          
\end{tabular}
}
\end{table}

% \begin{table*}
% \centering
% \caption{Quality metrics WIDE}\label{tab:quality}
% \begin{tabular}{c|ccc|ccc|}
% \toprule
% Method  & \multicolumn{3}{|c|}{Video1} & \multicolumn{3}{|c|}{Video2}\\
% &  PSNR& SSIM & LPIPS &  PSNR& SSIM & LPIPS\\
% \midrule
% HEVC & & & & & & \\
% AV1 & & & & & &\\
% $Ours_{low}$ & & & & & &\\
% $Ours_{high}$  & & & & & &\\
% \bottomrule
% \end{tabular}
% \begin{tabular}{c|ccc|ccc|}
% \toprule
% Method  & \multicolumn{3}{|c|}{Video1} & \multicolumn{3}{|c|}{Video2}\\
% &  PSNR& SSIM & LPIPS &  PSNR& SSIM & LPIPS\\
% \midrule
% HEVC & & & & & & \\
% AV1 & & & & & &\\
% $Ours_{low}$ & & & & & &\\
% $Ours_{high}$  & & & & & &\\
% \bottomrule
% \end{tabular}
% \end{table*}
% % multicolumn

The results regarding \textbf{computational time} are shown in Table~\ref{tab:speed}. 
Our proposed codec allows for an average increase in performance of 197\% compared to HEVC and AV1 and an increase of 91\% over the tiling technique. This increase is even more significant when the lower quality version of our codec is used. %, with an increase of up to XX\% in decoding speed.
In the experiment, the foveated decoding ($Ours_{FD}$) is applied on the wavelet-based video with high quality settings. Due to the foveation, the performance increases by 223\% over HEVC allowing a better performance than the lower quality wavelet-encoded videos.
Please note that we used the hardware accelerated on the GPU for the decoding of HEVC and AV1. The dedicated decoding chips allow for significant increases in decoding speed compared to conventional decoding. Additionally, the compute shaders for the mapping and rendering can be executed in parallel to the decoding through the dedicated chips.  
A comparable chip for decoding wavelet transforms could also significantly improve the performance of a wavelet-based codec while the compute unit can be used for other tasks.

%% Quality:
We compared the results' \textbf{quality} of all codecs by the commonly used metrics PSNR, SSIM~\cite{ssim}, and LPIPS~\cite{lpips}. The given values are averages over all frames and compared with the uncompressed reference frames (see Table~\ref{tab:quality}).
In terms of image quality our method performs equally to the other codecs, HEVC/H.265 and AV1, when high quality settings are chosen. As can be expected, the image quality is on a lower level when the low quality parameters are chosen for the wavelet-based encoding.  

%%% Compression:
The \textbf{compression rates} of the wavelet files in both quality configurations can be seen in Table~\ref{tab:compression}. With our wavelet-based approach, we are able to compress the raw information to over one hundredth in size for most videos.
%This significant data reduction allows the storage of hundreds of videos.
Compared to HEVC and AV1 compression we achieve about half the compression rates, depending on the quality of the video.
The tiled HEVC videos by the technique of Zare\etal~\cite{Zare:2016} are on average three times larger than our wavelet-compressed video files due to the significant compression losses of the tiling process.

\begin{table}
\centering
\caption{\textbf{Compression ratios} of the wavelet video files in relation to the uncompressed data.}
\label{tab:compression}
\centering
\setlength\tabcolsep{2 pt}%9
{\small%\footnotesize%\scriptsize%
\begin{tabular}{ccccccccc}
\toprule
              &City$_{CGI}$  &  Mountain$_{CGI}$ & Downhill & Horse  & Climbing & Walking & Cave* & Boat* \\
\midrule     
$Ours_{LQ}$ &   176:1      &    250:1         &  147:1   & 187:1  & 250:1    & 185:1   & 714:1 & 312:1 \\
$Ours_{HQ}$ &   52:1       &    75:1          &    77:1    & 100:1  & 128:1    & 100:1   & 416:1 & 117:1 \\
\bottomrule
\end{tabular}
}
\end{table}

%%%%%%%%%%%%%%%%%%%%%%%%%%%%%%%%%%%%%%%%%%%%%%%%%%%%%%%%%%%%%%%%
\subsection{Perceptual Evaluation} %subjective evaluation
%%%%%%%%%%%%%%%%%%%%%%%%%%%%%%%%%%%%%%%%%%%%%%%%%%%%%%%%%%%%%%%%

In a next step, we compare the codecs in terms of observer's preferences to reveal subtle perceptual differences that were not found by the objective metrics. 
In this perceptual experiment, we are particularly interested in analyzing to what extent the higher frame rates we can achieve with our wavelet codec also contribute to a better VR experience.
%%% codecs
As before, HEVC and AV1 serve as comparison techniques. The wavelet videos use the high quality version that showed to be most comparable to the other codecs in quality while still maintaining high display speeds.
%%% conditions
%Based on the results of the display speed measurement, a higher streaming rate of the relevant data can be achieved with a wavelet-based codec. 
Due to the better data utilization of a wavelet-based codec, a higher resolution of the video or a higher frame rate can be provided. 
Thus, there are two conditions to be considered: the videos are in the same resolution but have different frame rates (\textit{Speed} condition), or the videos are in different resolutions but provide the same frame rate (\textit{Quality} condition).
In the \textit{Speed} condition all videos are at a resolution of $8192\times8192$ pixels. While the comparison techniques provide common 30\ac{fps} in this condition, the wavelet videos provides double the frame rate (60\ac{fps}) based on the results of the display speed measurements.
In the \textit{Quality} condition the frame rate of all videos is set to 60\ac{fps} but the HEVC and AV1 videos, here, have a resolution of $4096\times4096$ pixels. The length of at videos is 10 seconds.
%%% scenes
For the perceptual evaluation we only use the \ac{cgi} scenes. The upscaled real-world videos have a corrupted stereo view because the upscaling algorithms are not designed for stereo footage. This display error should be irrelevant for the image-based metrics, but makes perceptual experiments impossible. 

%%% procedure
As the abstract feeling of comfort in a VR experience cannot be represented by a linear scale, we utilize the paired comparisons technique~\cite{Kendall:1940}.
Given the same video with two different encodings played immediately after each other, the participants were instructed to choose the video they would prefer for a presentation in \ac{vr}. 
The question was intentionally kept open and participants were free to base their decision on the visual quality, the temporal smoothness of the video, or a lower incidence of cybersickness. 

Given three codecs, there are three possible comparisons per video: Ours vs. AV1, Ours vs. HEVC, and AV1 vs. HEVC. This results in a total of $12$ decisions per person given two scenes for both conditions. In the experiment, the order of the paired stimuli comparisons for each participant was counterbalanced to avoid side effects. 

%%% participants and apparatus 
A total of $23$ participants took part in the experiment ($10$ females, age range = $22-59$, avg. age = $34.83$, SD = $13.04$) resulting in $184$ votes \textit{per codec}. 
The videos were shown in a HTC Vive Pro Eye \ac{hmd}.
For every comparison, the participants were ask to maintain a fixed head position to compare the same part of the 360\degree~video.

\subsubsection{Analysis and Results}

%Figure~\ref{fig:pairCompare} shows the results of the paired comparisons of the codecs. The methods are ranked by the number of votes they received, respectively the number of times a method was preferred over another.

On a first analysis, the voting of the participants ($n=23$) leads us to the displayed results in the top half of Table~\ref{tab:PercentNoConsistent} ("Raw Data", column "\% Preferences"). 
The results show the analysis for each scene as well as an analysis per condition combining both scenes (labelled as "AVG").
%thus, providing a maximum of votes per method of $46$,
%%%%%%%%%%%%%%%%%%%%%%%%%%%%%%%%%%%%%%%%%%%%%%%%%%%%%%%%%%%%%%%%%%%%%%%%%%%%%
%%%     TABLE AVG VALUES PER SCENE/CONDITION ALL PPANTS
% \usepackage{multirow}
\begin{table*}[th]
\centering
\caption{\textbf{Perceptual results.} The top half refers to the uncorrected data of all participants ($n=23$). In the bottom half the results are corrected for consistency ($n_{Speed}=21, n_{Quality}=15$).}
\label{tab:PercentNoConsistent}
    \small{
    \setlength\tabcolsep{10.5pt}
    \begin{tabular}{l|l|l|ccc|c|c|c|c}
         &  &  & \multicolumn{3}{c|}{\textbf{\% Preferences}} &  &  &  &  \\ %\cline{4-6}
    \multirow{-2}{*}{\textbf{ANALYSIS}} & \multirow{-2}{*}{\textbf{CONDITION}} & \multirow{-2}{*}{\textbf{SCENE}} & \multicolumn{1}{c|}{Ours vs. AV1} & \multicolumn{1}{c|}{Ours vs. HEVC} & AV1 vs. HEVC & \multirow{-2}{*}{\textbf{$\zeta$}} & \multirow{-2}{*}{\textbf{$u$}} & \multirow{-2}{*}{\textbf{$\chi^2$}} & \multirow{-2}{*}{\textbf{$p$}} \\ \hline\hline
    %%%%%%%%%%%%%%%%%%%% RAW DATA
        &  & City & \multicolumn{1}{c|}{\textbf{95.65\%}} & \multicolumn{1}{c|}{\textbf{82.61\%}} & 4.35\% & 0.96 & 0.68 & 48.13 & \textless .0001 \\
        & \textbf{Speed } & Mountain & \multicolumn{1}{c|}{\textbf{100.00\%}} & \multicolumn{1}{c|}{\textbf{56.52\%}} & 8.70\% & 0.96 & 0.55 & 39.09 & \textless .0001 \\
       &  & \cellcolor[HTML]{C0C0C0}\textbf{AVG} & \multicolumn{1}{c|}{\cellcolor[HTML]{C0C0C0}\textbf{97.83\%}} & \multicolumn{1}{c|}{\cellcolor[HTML]{C0C0C0}\textbf{69.57\%}} & \cellcolor[HTML]{C0C0C0}6.52\% & \cellcolor[HTML]{C0C0C0}0.96 & \cellcolor[HTML]{C0C0C0}0.60 & \cellcolor[HTML]{C0C0C0}84 & \cellcolor[HTML]{C0C0C0}\textless .0001 \\ \cline{2-10} 
        &  & City & \multicolumn{1}{c|}{\textbf{56.52\%}} & \multicolumn{1}{c|}{\textbf{56.52\%}} & 21.74\% & 0.87 & {\color[HTML]{CC4125} \textbf{0.08}} & 8.13 & {\color[HTML]{CC4125} \textbf{.0434}} \\
       & \textbf{Quality } & Mountain & \multicolumn{1}{c|}{\textbf{86.96\%}} & \multicolumn{1}{c|}{43.48\%} & 26.09\% & 0.74 & {\color[HTML]{CC4125} 0.23} & 18.22 & .0004 \\
        \multirow{-6}{*}{\textbf{Raw Data}} &  & \cellcolor[HTML]{C0C0C0}\textbf{AVG} & \multicolumn{1}{c|}{\cellcolor[HTML]{C0C0C0}\textbf{71.74\%}} & \multicolumn{1}{c|}{\cellcolor[HTML]{C0C0C0}\textbf{50.00\%}} & \cellcolor[HTML]{C0C0C0}23.91\% & \cellcolor[HTML]{C0C0C0}0.80 & \cellcolor[HTML]{C0C0C0}\textbf{\color[HTML]{CC4125} 0.13} & \cellcolor[HTML]{C0C0C0}20.55 & \cellcolor[HTML]{C0C0C0}{\color[HTML]{000000} .0001} \\ \hline\hline
    %%%%%%%%%%%%%%%%%%%% CORRECTED DATA
       &  & City & \multicolumn{1}{c|}{\textbf{100.00\%}} & \multicolumn{1}{c|}{\textbf{80.95\%}} & 4.76\% & 1 & 0.72 & 46.24 & \textless .0001 \\
        & \textbf{Speed } & Mountain & \multicolumn{1}{c|}{\textbf{100.00\%}} & \multicolumn{1}{c|}{\textbf{57.14\%}} & 4.76\% & 1 & 0.59 & 38.62 & \textless .0001 \\
        &  & \cellcolor[HTML]{9AFF99}\textbf{AVG} & \multicolumn{1}{c|}{\cellcolor[HTML]{9AFF99}\textbf{100.00\%}} & \multicolumn{1}{c|}{\cellcolor[HTML]{9AFF99}\textbf{69.05\%}} & \cellcolor[HTML]{9AFF99}4.76\% & \cellcolor[HTML]{9AFF99}1 & \cellcolor[HTML]{9AFF99}0.65 & \cellcolor[HTML]{9AFF99}82.48 & \cellcolor[HTML]{9AFF99}{\color[HTML]{000000} \textless .0001} \\ \cline{2-10} 
       &  & City & \multicolumn{1}{c|}{\textbf{80.00\%}} & \multicolumn{1}{c|}{\textbf{60.00\%}} & 20.00\% & 1 & {\color[HTML]{036400} \textbf{0.20}} & 11.4 & {\color[HTML]{036400} .0097} \\
        & \textbf{Quality } & Mountain & \multicolumn{1}{c|}{\textbf{86.67\%}} & \multicolumn{1}{c|}{46.67\%} & 6.67\% & 1 & {\color[HTML]{036400} 0.39} & 19.4 & .0002 \\
       \multirow{-6}{*}{\textbf{\begin{tabular}[c]{@{}l@{}}Corrected \\ for \\ Consistency\end{tabular}}} &  & \cellcolor[HTML]{9AFF99}\textbf{AVG} & \multicolumn{1}{c|}{\cellcolor[HTML]{9AFF99}\textbf{83.33\%}} & \multicolumn{1}{c|}{\cellcolor[HTML]{9AFF99}\textbf{53.33\%}} & \cellcolor[HTML]{9AFF99}13.33\% & \cellcolor[HTML]{9AFF99}1 & \cellcolor[HTML]{9AFF99}\textbf{\color[HTML]{036400} 0.31} & \cellcolor[HTML]{9AFF99}29.60 & \cellcolor[HTML]{9AFF99}{\color[HTML]{000000} \textless .0001}
\end{tabular}
} % end \small
\end{table*}

%%%%%%%%%%%%%%%%%%%%%%%%%%%%%%%%%%%%%%%%%%%%%%%%%%%%%%%%%%%%%%%%%%%%%%%%%%%%%

In order to assess the results of the paired comparisons we follow the methodology from Setyawan and Lagendijk~\cite{Setyawan:2004}. We first study the consistency of choices within one participant as well as the agreement in choices among all participants.
%%% Coefficient of consistency 
The coherence of the answers of the participants is indicated by the \textbf{coefficient~of~consistency} $\zeta \in [0,1]$, with $\zeta=1$ implying perfect consistency.
Low consistency values of single participants can indicate that these individuals had difficulties to differentiate between the stimuli and thus, we can expect their judgement abilities to be worse than the average.
As the number of methods $m=3$, $\zeta<1$ can only arise with the occurrence of one kind of circular triad such that $C_1 \rightarrow C_2 \rightarrow C_3$ but $C_3 \rightarrow C_1$. 
In Table~\ref{tab:PercentNoConsistent} we present the coefficient of consistency as an average over all participants.
%%% Coefficient of agreement
The consistency of choices of single participants does not necessarily mean that identical choices were made between all participants. The diversity of preferences for the number of participants $n$ is described by the \textbf{coefficient of agreement} $u$. Complete agreement is achieved with  $u=1$, meaning that all participants favoured the same method for all decisions.
High disagreement indicated by low $u$ values, on the other hand, suggests that participants had difficulties to make a joint choice. 
Disagreement may suggests either that the stimuli were perceived as not distinguishable or a general split of opinions about the stimuli. Here, the minimum $u$, and accordingly complete disagreement, is given by $u_{min}=-\frac{1}{(n-1)} \approx -0.045$.
By the definition of Kendall and Babington-Smith~\cite{Kendall:1940}, the coefficient of agreement $u$ is derived by
\begin{equation}
    u = \frac{2\tau}{\binom{n}{2}\binom{m}{2}}, \text{where} \tau = \sum_{i=1}^m \sum_{j=1}^m \binom{a_{ij}}{2} 
\end{equation}
with $a_{ij}$ is the number of times method $i$ is chosen over method $j$, whereby $i\neq j$. As before, the number of participants is denoted by $n$ and the number of methods by $m$.

We determine the statistical significance of $u$ by testing against the null hypothesis that all votes  were chosen randomly. For a significant $u$ we can conclude the alternative hypothesis that the agreement is above the value one would expect from random choices.
To determine the significance under the null hypothesis we perform a chi-squared test ($\chi^2$). 
As proposed by former research~\cite{Siegel:1988}, with our $n$ we can derive $\chi^2$ in simple form. 
\begin{equation}
    \chi^2 = \binom{m}{2} [1+u(n-1)]
\end{equation}
Since we compare three codecs, the $\chi^2$ distribution is evaluated with $\binom{m}{2}=3$ \ac{dof}.
Therefore, the statistical significance at level $p$ derives by $\chi^2_3$.

%%%%%%%%%%%%%%%%%%%%%%%%%%%%%%%%%%%%%%%%%%%%%%%%%%

%While the values are acceptable, one can see that, while the average consistency is high, the coefficient of agreement in the Quality condition drops in relation to the Speed condition (numbers highlighted in red in Table~\ref{tab:PercentNoConsistent}), sinking to a really low agreement for the City scene, where results are not anymore statistically significant for our targeted threshold of $\alpha=0.99$. 
The analysis of the raw data for consistency and agreement is shown in Table~\ref{tab:PercentNoConsistent}.  
While the participants show a high consistency, the coefficient of consistency $\zeta$ for the \textit{Quality} condition is considerably lower than for the \textit{Speed} condition. The difference in agreement is even more pronounced and the participants show clear disagreement in the ratings of \textit{Quality} (numbers highlighted in red). This mismatch is most apparent in the city scene where results are not statistically significant anymore for the targeted threshold of $p<.01$.

%A deeper look into the consistency of each individual participant while making their decisions further explains this. 
A more in-depth analysis of the consistency of individual ratings shows the cause for this outcome. 
In the \textit{Quality} condition, a total of eight participants were not able to properly distinguish between stimuli, leading them to produce triads. Also, two participants had difficulties emitting judgement in the \textit{Speed} condition.
All the rest of our participants show perfect judgement capability ($\zeta=1$, no triads), indicating that the inconsistencies arised from the individual participant's ability to judge and not from a problem with a consensus between participants. 
Therefore, we correct the analysis for a consistent result and remove the votes for the entire condition of all those participants who did not show perfect consistency. %Subtracting these individuals from the corresponding conditions 
By rerunning the analysis with the votings corrected for perfect consistency ($n_{Speed}= 21, n_{Quality}= 15$) we can obtain the values corresponding to participants with perfect discerning abilities. These results are show in the bottom half of Table~\ref{tab:PercentNoConsistent} ("Corrected for Consistency"). 
While the \textit{Quality} condition still seems to be more controversial, the agreement of the participants for this condition increases substantially (numbers highlighted in green).
With the correction, all differences in preference are statistically significant for $p<0.01$. 

In general, the results of the perceptual study indicated that the videos using both our codec and HEVC were favored over AV1 for all conditions. 
The conclusions of the comparison between our codec and HEVC are dependent on the analyzed condition. While the participants clearly favoured our codec in the \textit{Speed} condition, the perceived quality overall is on par with the H.265 compressed video. A more in-depth analysis reveals the choice of scene as a decisive factor for the participants' preference. 
Straight lines and geometric structures are key attributes of urban scenery. In these representations with man-made content, quality differences and artifacts seem to be more conspicuous. 
The \textit{Mountain} scene, on the other hand, is a natural scene and, as such, its content is less structured and highly semantically homogeneous due to the recurring textures. In this scene, the participants showed difficulties to distinguish the quality of the different compression techniques. These observations are consistent with the findings of previous research~\cite{Rubinstein:2010, Castillo:2011}.
In the \textit{Speed} condition, the scene attributes seem to became even more salient for the observers' choice, drifting their preferences towards our method (preference $>80\%$).

%\SUS{Short discussion on "A deeper look into the participants' preferences when deciding between favouring W or H could be due to the scenario/image attributes. City portraits a hand made scenario, where straight lines and geometric structures are key attributes of the scene, while Mountains is a natural scene with its content is obviously less structured and highly semantically homogeneous due to the recurring textures. In Speed condition this seems to became even more salient for the observers drifting their preferences towards our method. These observation is consistent with the findings of previous research~\cite{Rubinstein:2010, Castillo:2011} }
% (https://people.csail.mit.edu/mrub/retargetme/) 

% %%%%%
% Need to do now the Ranking, \textbf{W(3, 0.01) = 4.125, Setwayan 2.2.3} we expect in both analysis same groupings:\\
% Raw Data: Speed : (W) (H) (A)\\
% Quality: (W H) (A)\\

% Corrected Data: Speed : (W) (H) (A)\\
% Quality: (W H) (A)\\

\section{Discussion and Limitations}

In the following, we discuss further key points of consideration and address limitations of the current implementation

%% experimental results
\textbf{Experimental Results.}
We compared our wavelet-based codec against two common video codecs and previous work.
For the evaluation, we considered a low and high quality version of the wavelet-encoded videos, because in a practical application either quality or speed may be prioritised.
The results show that the codec can be optimised for such requirements by changing the encoding parameters.
However, even at the highest quality, we achieve significantly higher decoding speeds than the other methods.
%For the foveated decoding, the quality in the foveal area corresponded to the higher resolution version of the video.
The foveated decoding technique leverages the properties of the human visual system, resulting in peripheral resolution differences that are unnoticeable to VR users compared to fully resolved viewports~\cite{Leigh:2015}.
%The visual quality of the frame in the periphery exceeds that of classic foveated rendering, because a transformation step is performed instead of upscaling (see Fig.\ref{}).
At the same time, the foveated decoding allows for the highest decoding speeds.
In a perceptual experiment, we studied the importance of the frame rate and video resolution for the perceived quality of the VR experience. The results suggest that for \vids in VR, the frame rate is of significantly greater importance to users than the image quality. This emphasis on frame rate for the user experience is consistent with former research~\cite{Banitalebi:2015}.

\textbf{Single Operation Point.}
%\subsection{Major Application Scenario}
Our codec is designed for a very specific use-case.
The core motivation is to have 360° videos with a resolution and display speed that does not induce cybersickness and is pleasant to watch.
In our investigations, we explored in detail the single operation point that best covers this scenario. 
%For the experiment we fine-tuned the comparison methods to have the best performance at a quality that is still reasonable for VR, whereby file size was not a priority.
For a fair comparison we chose the parameters of our codec so that the quality is on the same level. With this baseline, we then measured the speed of the methods.
A broader range of quality-rate scenarios can be explored in the future to utilize wavelet-based coding for a verity of applications.

\textbf{Streaming.}
So far, we have primarily addressed videos that are stored on a local drive. Online streaming is another common way to retrieve video data.
With online streaming, the amount of data that is transmitted is much more relevant due to bandwidth limitations. For these limitations, a wavelet-based codec benefits from the direct viewport-dependent streaming from file. This property allows to reduce the transfer rates by up to ten times compared to the total size of the video. In its current form, our encoding is not yet optimized for real time execution, which we plan to address as a natural next step.

\textbf{File Size.}
Our wavelet format does not use any container format but is stored in simple binary form. Neither is a color transformation performed, for example to the YUV space. Such techniques are applied by other codecs to reduce their file sizes to the minimum while preserving the best possible quality.
In this paper the major focus was on display speed.
In future work such techniques may be introduced to further reduce the file sizes of wavelet-based video coding.

\textbf{Reference Data.}
%\subsection{Limitations}
Our objective with the reference data scaling of the real-world videos was to generate uncompressed high-resolution, high frame-rate video data. We used a combination of downscaling followed by AI-based upscaling to remove compression artifacts from the original videos. This removal is not perfect and it can be assumed that the compression rate of a wavelet-based codec is significantly higher for raw footage.
Such a use of a wavelet-based codec can only be achieved when the encoding is directly performed by the capturing device with the native color information.

\textbf{Professional Filming.}
The videos from our experiment are considered as casual recordings.
Nevertheless, \vids are not only used by amateurs, but also by professional filmmakers. For professional filming, it can be necessary to display different areas of a frame in different qualities, such as the background or the masks of an actor, which stands out as artificial in high resolutions.
With conventional methods, this procedure requires post-processing or recapturing of the video. With a wavelet-based codec a pre-adjustment is not necessary and the video can be stored in full resolution. Individual quality levels may be chosen at decoding time for defined parts of the video, comparable to our foveated decoding approach (cf.~Sec.~\ref{sec:foveation}).

\textbf{Eyetracking.}
In VR, eye tracking is nowadays mostly used for computer-generated content, where foveation allows for significant increases in rendering speed.
The foveated decoding of our codec opens up the opportunity for an broader use of eye tracking in VR where it can be used to increase the playback speed of \vids through unobtrusive quality gradation in the peripheral area.  

%\cg{This is new:}
\textbf{Wide \Ac{fov}.}
When the \ac{fov} gets unusually wide, this would affect the performance of our approach since the decoding is viewport dependent. The headset with the widest \ac{fov} currently on the market is the \textit{StarVR One} with an overall horizontal \ac{fov} of 210\degree~and a vertical \ac{fov} of 130\degree~\cite{StarVR}.
Exploring this scenario, we found that with wavelet coding we are still able to achieve $>100$ \ac{fps} with the high quality configuration in all scenes. With foveated decoding applied, the frame rate is significantly higher.
In comparison, the tiling approach with this wide \ac{fov} no longer yields any performance benefit and actually performs worse than the native full frame HEVC decoding ($\approx50$ \ac{fps}).

% %% why wavelet videos are unpopular
% Fakt ist jedoch, dass wavelet basierte codecs für bilder und videos bis heute keine verbreitung gefunden haben. Die gründe können vielseitig sein, werden aber oft darauf zurückgeführt, dass ihr einsatzgebiet bis jetzt immer auf die klasischen einsatzgebiete von video display abziehlten, für die die klasischen codecs allerdings schon hochgradig perfektioniert sind.

% %% empathise for wavelet use  
% 360° videos könnten ein einsatzgebiet darstellen, für das wavelet codecs noch einmal sehr interessant werden könnten, da dort ihre eigentlichen vorteile tatsächlich zum tragen kommen.

\section{Conclusion}

In this paper we proposed wavelet-based video coding for fast and high-resolution playback of 360\degree~videos.
%
% We showed that with a wavelet based compression approach, videos can be considered \SUS{flexible by arbitrary regions} which in the case of \vids is key for a fast decoding.
%
We showed that our wavelet-based compression approach allows for selective loading and decoding of arbitrary video regions, which in the case of \vids is key for a fast decoding.
While in our experiment our codec reached display speeds at least two times higher than the other methods tested, the quality remained at a comparable level.
The importance of high frame rates for a good VR experience is supported by the results of our perceptual experiment.
In addition, with our codec we have introduced foveated decoding, allowing for an unobtrusive quality decrease in the outer regions of the view. Foveated decoding can be applied on run-time and further increases the decoding times.
In conclusion, wavelet-based video approaches solve the problems that are raised by \ac{dct} codecs when a fast or viewport-dependent playback of \vids is required. Especially for VR environments, wavelet-based codecs show to be a valuable extension, offering the opportunity to display \vids in a quality and speed comparable to renderings of virtual worlds.

\bibliographystyle{abbrv-doi}

\bibliography{allReferences}

\end{document}